\documentclass{ws-procs9x6}

\begin{document}

\title{Dropping $\rho$ and $A_1$ Meson Masses\\
at the Chiral Phase Transition\\
in the Generalized Hidden Local Symmetry\footnote{%
Talk presented by M.~Harada at 2006 International Workshop on 
``Origin of Mass and Strong Coupling Gauge Theories'' (SCGT06)
based on the work done in Ref.~\refcite{HS:GHLS}.
}}

\author{Masayasu Harada} 

\address{
Department of Physics, Nagoya University,
Nagoya, 464-8602, Japan\\
E-mail: harada@hken.phys.nagoya-u.ac.jp}

\author{Chihiro Sasaki}

\address{
Gesellschaft f\"{u}r Schwerionenforschung (GSI),
64291 Darmstadt, Germany\\
E-mail: c.sasaki@gsi.de}

\begin{abstract}
In this contribution to the proceedings, 
we present our recent analysis on
the chiral symmetry restoration in the generalized
hidden local symmetry (GHLS) which incorporates the
rho meson, $A_1$ meson and the pion
consistently with the chiral symmetry of QCD.
It is shown that
a set of parameter relations, which ensures the first and second 
Weinberg sum rules, is invariant under the renormalization
group evolution.
Then, it is found that the Weinberg sum rules
together with the
matching of the vector and axial-vector current correlators
to the OPE
inevitably leads to {\it the dropping masses of both rho and 
$A_1$ mesons} at the symmetry restoration point,
and that the mass ratio 
flows into one of three fixed points.
\end{abstract}


\bodymatter

\section{Introduction}\label{sec:intro}

Changes of the hadron masses are indications
of the chiral symmetry restoration 
occurring in hot and/or dense QCD~\cite{rest}.
Dropping masses of hadrons following the Brown-Rho (BR)
scaling~\cite{BR-scaling} can be 
one of the most prominent candidates of the strong
signal of the melting quark condensate 
$\langle\bar{q}q\rangle$ which is
the order parameter of the spontaneous chiral symmetry
breaking.

The vector manifestation (VM)~\cite{HY:VM} is the
Wigner realization in which
the $\rho$ meson becomes massless degenerate with the
pion at the chiral phase transition point.
The VM is formulated~\cite{HY:PRep,HS:VM,HKR:VM} 
in the effective field theory (EFT) based
on the hidden local symmetry (HLS)~\cite{BKUYY,BKY:PRep}.
The VM gives
a theoretical description of the dropping $\rho$ mass,
which is protected by the existence of the
fixed point (VM fixed point).

The dropping mass is supported by
the enhancement of dielectron mass spectra below
the $\rho / \omega$ resonance observed at CERN SPS~\cite{ceres,LKB},
the mass shift of the $\omega$ meson in nuclei measured by
the KEK-PS E325 Experiment~\cite{KEK-PS},
the CBELSA/TAPS Collaboration~\cite{trnka}
and also that of
the $\rho$ meson observed in the STAR experiment~\cite{SB:STAR}.
Recently NA60 Collaboration has provided data for the dimuon 
spectrum~\cite{NA60} and it seems difficult to explain the data
by a naive dropping $\rho$~\cite{NA60:2}.
However, there are still several ambiguities which were not
considered~\cite{Brown:2005ka-kb,HR,SG}.
Especially, the strong violation of the vector dominance (VD),
which is one of the significant predictions of the VM~\cite{HS:VD},
plays an important role~\cite{Brown:2005ka-kb,HS:dilepton} 
to explain the data.

In the VM, it was assumed that the axial-vector and scalar mesons
are decoupled from the theory near the phase transition point.
However, the masses of these mesons may decrease following the
BR scaling~\cite{BLR}.
There were several analyses with models including 
axial-vector mesons such as in Ref.~\refcite{rhoA1model}.
These analyses are not based on the fixed point structure
and found no significant reduction of the masses of
axial-vector meson.
Then, it is desirable to construct an EFT
which includes the axial-vector meson as a dynamical degree of
freedom, and to study 
whether a fixed point structure exists and it can realize
the light axial-vector meson.

In Ref.~\refcite{HS:GHLS}, 
we picked up 
the model based on the generalized hidden local symmetry
(GHLS)~\cite{BKY:NPB,BFY:GHLS,BKY:PRep,Kaiser:1990yf},
and developed the chiral perturbation theory
(ChPT) with GHLS.
We showed that
a set of the parameter relations,
which satisfies the pole saturated forms of the first and second
Weinberg sum rules,
is stable against the renormalization group evolution.
Then, we found that
the Weinberg sum rules together with 
the matching to the operator product expansion
necessarily leads to the dropping 
masses of both vector and axial-vector mesons.
Interestingly, the ratio of masses of vector and axial-vector 
mesons as well as the mixing between the pseudoscalar and
axial-vector mesons flows into one of three fixed points:
They exhibit the VM--like, Ginzburg-Landau--like 
and Hybrid--like patterns of the chiral symmetry restoration.

In the following, we briefly summarize the work done in 
Ref.~\refcite{HS:GHLS}.

\section{Generalized Hidden Local Symmetry}
\label{sec:GHLS}

The Lagrangian 
of the generalized hidden local symmetry 
(GHLS)~\cite{BKY:NPB,BKY:PRep,BFY:GHLS}
is based on
the $G_{\rm{global}} \times G_{\rm{local}}$ symmetry,
where $G_{\rm global}=[SU(N_f)_L \times SU(N_f)_R]_{\rm global}$ 
is the chiral symmetry and 
$G_{\rm local}=[SU(N_f)_L \times SU(N_f)_R]_{\rm local}$ 
is the GHLS.
The whole symmetry $G_{\rm global}\times G_{\rm local}$
is spontaneously broken to a diagonal $SU(N_f)_V$.
The basic quantities are
the GHLS gauge bosons $L_\mu$ and $R_\mu$,
which are identified with the vector and axial-vector mesons as
$V_\mu = (R_\mu + L_\mu)/2$ and 
$A_\mu = (R_\mu - L_\mu)/2$,
and 
three matrix valued variables $\xi_L$, $\xi_R$
and $\xi_M$, which are introduced as
$U = \xi_L^\dagger \xi_M \xi_R$.
Here $N_f \times N_f$ special-unitary matrix $U$ is
a basic ingredient of the chiral perturbation theory 
(ChPT)~\cite{ChPT}.

The fundamental objects are the Maurer-Cartan 1-forms
defined by
\begin{eqnarray}
&&
\hat{\alpha}_{L,R}^\mu = D^\mu\xi_{L,R}\cdot\xi_{L,R}^\dagger /i\,,
\quad
\hat{\alpha}_M^\mu = D^\mu\xi_M\cdot\xi_M^\dagger /(2i)\,,
\end{eqnarray}
where
the covariant derivatives of $\xi_{L,R,M}$ are given by
\begin{eqnarray}
&&
D_\mu \xi_L 
 = \partial_\mu\xi_L - iL_\mu\xi_L + i\xi_L{\cal{L}}_\mu\,,
\quad
D_\mu \xi_R 
 = \partial_\mu\xi_R - iR_\mu\xi_R + i\xi_R{\cal{R}}_\mu\,,
\nonumber\\
&&
D_\mu \xi_M 
 = \partial_\mu\xi_M - iL_\mu\xi_M + i\xi_M R_\mu\,,
\end{eqnarray}
with ${\cal{L}}_\mu$ and ${\cal{R}}_\mu$ being the external
gauge fields introduced by gauging $G_{\rm{global}}$.
There are four independent terms with the lowest derivatives:
\begin{eqnarray}
&&
{\cal L}_V 
 = F^2 \mbox{tr}\bigl[ \hat{\alpha}_{\parallel\mu}
   \hat{\alpha}_\parallel^\mu \bigr]\,,
\quad
{\cal L}_A 
 = F^2 \mbox{tr}\bigl[ \hat{\alpha}_{\perp\mu}
   \hat{\alpha}_\perp^\mu \bigr]\,,
\quad
{\cal L}_M 
 = F^2 \mbox{tr}\bigl[ \hat{\alpha}_{M\mu}
   \hat{\alpha}_M^\mu \bigr]\,,
\nonumber\\
&&
{\cal L}_\pi 
 = F^2 \mbox{tr}\bigl[ \bigl( \hat{\alpha}_{\perp\mu}
   {}+ \hat{\alpha}_{M\mu} \bigr)
   \bigl( \hat{\alpha}_{\perp}^\mu
   {}+ \hat{\alpha}_{M}^\mu \bigr)\bigr]\,,
\label{lag a-d}
\end{eqnarray}
where $F$ is the parameter
carrying the mass dimension $1$
and 
$
\hat{\alpha}_{\parallel,\perp}^\mu
 = \bigl( \xi_M\hat{\alpha}_R^\mu\xi_M^\dagger 
  {}\pm \hat{\alpha}_L^\mu \bigr)/2\,$.
The kinetic term of the gauge bosons 
is given by 
\begin{equation}
{\cal L}_{\rm kin}(L_\mu,R_\mu)
 = {}- \frac{1}{4g^2}\mbox{tr}\bigl[ L_{\mu\nu}L^{\mu\nu}
   {}+ R_{\mu\nu}R^{\mu\nu} \bigr]\,,
\label{lag kin}
\end{equation}
where $g$ is the GHLS gauge coupling 
and the field strengths are defined by
$L_{\mu\nu}
 = \partial_\mu L_\nu - \partial_\nu L_\mu
  {}- i\bigl[ L_\mu, L_\nu \bigr]\,$ and
$R_{\mu\nu}
 = \partial_\mu R_\nu - \partial_\nu R_\mu
  {}- i\bigl[ R_\mu, R_\nu \bigr]\,$.

By combining the above terms,
the GHLS Lagrangian is given by
\begin{equation}
{\cal L} = a{\cal L}_V + b{\cal L}_A + c{\cal L}_M
 {}+ d{\cal L}_\pi 
 {}+ {\cal L}_{\rm kin}(L_\mu,R_\mu)\,,
\label{lag p^2}
\end{equation}
where $a$, $b$, $c$ and $d$ are dimensionless parameters. 
{}From this
we find the following expressions for
the masses of vector and axial-vector mesons $M_{\rho,A_1}$,
the $\rho$-$\gamma$ mixing strength $g_{\rho}$
and
strength of the coupling of the $A_1$ meson to the axial-vector
current $g_{A_1}$:
\begin{eqnarray}
M_\rho = g \sqrt{a} F \,,
\quad
M_{A_1} = g \sqrt{b+c} F \,,
\quad
g_\rho = g a F^2 \,,
\quad
g_{A_1} = g bF^2 \,.
\label{g_rhopipi tree}
\end{eqnarray}

\section{Weinberg's Sum Rules}
\label{sec:WSR}

Let us start with 
the axial-vector and vector current correlators defined by
\begin{eqnarray}
&&
G_A(Q^2)(q^\mu q^\nu - q^2 g^{\mu\nu}) \delta_{ab}
=
\int d^4x\,e^{iqx}
\left\langle 0 \vert \, T\,
  J_{5a}^\mu(x) J_{5b}^\nu(0)
\vert 0 \right\rangle\,,
\nonumber\\
&&
G_V(Q^2)(q^\mu q^\nu - q^2g^{\mu\nu}) \delta_{ab}
=
\int d^4x\,e^{iqx}
\left\langle 0 \vert \, T\,
  J_a^\mu(x) J_b^\nu(0)
\vert 0 \right\rangle\,,
\end{eqnarray}
where $Q^2 = - q^2$ is the space-like momentum,
$J_{5a}^\mu$ and $J_a^\mu$ are the axial-vector and 
vector currents and $(a,b)=1,\ldots,N_f^2-1$.
At the leading order of the GHLS
the current correlators $G_{A,V}$ 
are expressed as
\begin{eqnarray}
G_A(Q^2)
= \frac{F_\pi^2}{Q^2} + \frac{F_{A_1}^2}{M_{A_1}^2+Q^2}\,,
\quad
G_V(Q^2)
= \frac{F_\rho^2}{M_\rho^2 + Q^2}\,,
\label{pole}
\end{eqnarray}
where the $A_1$ and $\rho$ decay constants are defined by
\begin{eqnarray}
F_{A_1}^2 = \Bigl( \frac{g_{A_1}}{M_{A_1}} \Bigr)^2 
= \frac{b^2}{b+c}F^2\,,
\quad
F_\rho^2 = \Bigl( \frac{g_\rho}{M_\rho} \Bigr)^2 
= a F^2\,.
\end{eqnarray}
The same correlators can be evaluated by the OPE~\cite{SVZ},
which shows that
the difference between two correlators scales as $1/Q^6$:
\begin{equation}
G_A^{\rm(OPE)}(Q^2) - G_V^{\rm(OPE)}(Q^2) 
= \frac{32\pi}{9} \frac{\alpha_s \,\langle \bar{q}q\rangle^2}{Q^6}
\ .
\label{A-V:OPE}
\end{equation}

We require that the high energy behavior of the difference between
two correlators in the GHLS agrees with that in the OPE:
$G_A(Q^2) - G_V(Q^2)$ in the GHLS scales as $1/Q^6$.
This requirement can be satisfied only if the following relations
are satisfied:
\begin{eqnarray}
F_\pi^2 + F_{A_1}^2 = F_\rho^2\,,
\quad
F_{A_1}^2 M_{A_1}^2 = F_\rho^2 M_\rho^2\,,
\label{WSR}
\end{eqnarray}
which are nothing but the pole saturated forms of the 
Weinberg first and second sum rules~\cite{Weinberg}.
In terms of the parameters of the GHLS Lagrangian,
the above relations can be satisfied
if we take 
\begin{equation}
a = b \,, \quad d = 0 \,.
\label{tsl}
\end{equation}

In Ref.~\refcite{HS:GHLS},
we calculated the RGEs for the parameters $a$, $b$, $c$, $d$ and 
the gauge coupling $g$.
It was shown that
the parameter relations $a = b$ and $d=0$ are stable
against the renormalization group evolution,
i.e., {\it the non-renormalization of the Weinberg sum rules
expressed in terms of the leading order parameters
in the GHLS}.

\section{Chiral Symmetry Restoration}
\label{sec:PS}

The maching condition together
with the Weinberg sum rules (\ref{WSR}) 
in the high-energy region is obtained as
\begin{eqnarray}
F_{A_1}^2 M_{A_1}^4 - F_\rho^2 M_\rho^4
=
F_\rho^2 M_\rho^2 \left( M_{A_1}^2 - M_\rho^2 \right) 
=
\frac{32\pi\alpha_s}{9} \langle \bar{q}q \rangle^2\,,
\label{match rest}
\end{eqnarray}
which is a measure of the spontaneous chiral symmetry breaking.
When the chiral restoration point is approached,
the quark condensate approaches zero:
$\langle \bar{q} q \rangle \ \rightarrow\  0 $.
Then, the 
parameters of the GHLS Lagrangian
scales as 
$a^2 \cdot c \cdot g^4 \propto \langle \bar{q}q \rangle^2
\ \rightarrow\ 0$
near the chiral symmetry restoration point.
Since $a=0$ and $c=0$ are unstable against the RGEs
and $g=0$ is a fixed point,
the symmetry restoration in the GHLS 
can be realized only if 
\begin{equation}
 g \ \rightarrow\ g^\ast = 0 \,.
\label{cond}
\end{equation}
This implies the massless $\rho$ and $A_1$ mesons,
since both masses are proportional to the gauge coupling
$g$.
Thus we conclude that,
{\it when we require the first and second Weinberg sum rules to be
satisfied,
the chiral symmetry restoration in the GHLS 
required through the matching to QCD 
can be realized with masses of $\rho$ and $A_1$ mesons vanishing
at the restoration point}:
\begin{equation}
M_\rho \ \rightarrow\ 0 \,, \quad
M_{A_1}\ \rightarrow\ 0 \,.
\end{equation}

To study the phase structure of the GHLS through the RGEs for
$a$, $c$ and $g$, it is convenient to
introduce the following dimensionless parameters
associated with $a, c$ and $g$:
\begin{eqnarray}
W(\mu) 
 = \frac{N_f}{2(4\pi)^2}\frac{a+c}{a c}\frac{\mu^2}{F^2} 
= \frac{N_f}{2(4\pi)^2}\frac{\mu^2}{F_\pi^2(\mu)}\,,
\quad
\zeta(\mu) = \frac{a}{a+c} = \frac{M_\rho^2}{M_{A_1}^2}\,.
\label{def:W zeta}
\end{eqnarray}
The phase of the GHLS is determined by the on-shell pion 
decay constant $F_\pi(\mu = 0)$,
or equivalently $W$ defined in Eq.~(\ref{def:W zeta}), 
as
\begin{eqnarray}
W(\mu=0)=0 \ : \ \mbox{broken phase}\ ,
\qquad
W(\mu=0)\neq 0 \ : \ \mbox{symmetric phase}\ .
\nonumber
\end{eqnarray}
\begin{figure}
 \begin{center}
  \includegraphics[width = 6.8cm]{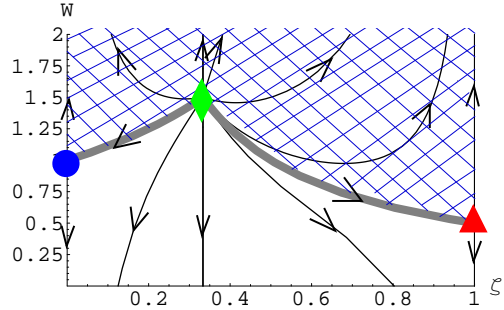}
 \end{center}
 \caption[]{Phase diagram on $\zeta$-$W$ plane. Arrows on the flows are
  written from the ultraviolet to the infrared. Gray lines divide
  the broken phase (lower side) and the symmetric phase 
  (upper side; cross-hatched area).
  Points denoted by $\blacktriangle$, $\bullet$ and $\blacklozenge$ 
  express the fixed point $(\zeta,W)=(1,1/2)$, $(0,1)$ and $(1/3,3/2)$
  respectively.}
 \label{fig:zW}
\end{figure}
The flow diagram shown in Fig.~\ref{fig:zW} has 
three fixed points:
\begin{eqnarray}
&&\mbox{GL-type :}\quad
(\zeta^\ast,W^\ast)=(1,1/2)\,,
\nonumber\\
&&\mbox{VM-type :}\quad
(\zeta^\ast,W^\ast)=(0,1)\,,
\nonumber\\
&&\mbox{Hybrid-type :}\quad
(\zeta^\ast,W^\ast)=(1/3,3/2)\,.
\label{FP-zW}
\end{eqnarray}
{}From this we can distinguish three patterns of the
chiral symmetry restoration characterized by three fixed points
by the values of the ratio of $\rho$ and $A_1$ meson masses
expressed by $\zeta$ as in Eq.~(\ref{def:W zeta}) as follows:
At the fixed point ``GL-type'',
$\zeta$ goes to 1, which implies that the $\rho$ meson mass
degenerates into the $A_1$ meson mass.
At the fixed point ``VM-type'', on the other hand,
the $\rho$ meson becomes massless faster than the $A_1$ meson
since $\zeta$ goes to zero.
The fixed point ``Hybrid-type'' 
is the ultraviolet fixed point in any direction,
so that it is not so stable as to ``GL-type'' and ``VM-type''.

To summarize, we found that
the chiral symmetry restoration in the GHLS 
required through the matching to QCD can be realized 
only if the masses of $\rho$ and $A_1$ mesons vanish
at the restoration point:
\begin{equation}
M_\rho \ \rightarrow\ 0 \,, \quad
M_{A_1}\ \rightarrow\ 0 \,,
\end{equation}
and that
the mass ratio flows into one of the 
following three fixed points:
\begin{eqnarray}
&&
\mbox{GL-type} \ : \ 
  M_\rho^2/M_{A_1}^2 \ \rightarrow\ 1 \,,
\nonumber\\
&&
\mbox{VM-type} \ : \ 
  M_\rho^2/M_{A_1}^2 \ \rightarrow\ 0 \,,
\nonumber\\
&&
\mbox{Hybrid-type} \ : \ 
  M_\rho^2/M_{A_1}^2 \ \rightarrow\ 1/3 \,.
\end{eqnarray}
In Ref.~\refcite{HS:GHLS} we studied the strength of 
the direct $\gamma$-$\pi$-$\pi$ coupling, which measures
the validity of the vector dominace (VD).
We have shown that,
$g_{\gamma\pi\pi} \ \rightarrow\ 0$ for the GL-type,
$g_{\gamma\pi\pi} \ \rightarrow\ 1/2$ for the VM-type
and 
$g_{\gamma\pi\pi} \ \rightarrow\ 1/3$ for the Hybrid-type.
This strongly affects to the understanding of the experimental data
on dilepton productions based on the dropping $\rho$.

\vspace{-0.3cm}

\section*{Acknowledgment}

The work of M.H. 
is supported in part by the Daiko Foundation \#9099, 
the 21st Century
COE Program of Nagoya University provided by Japan Society for the
Promotion of Science (15COEG01), and the JSPS Grant-in-Aid for
Scientific Research (c) (2) 16540241.
The work of C.S. is supported in part by the Virtual Institute
of the Helmholtz Association under the grant No. VH-VI-041.


\begin{thebibliography}{99}
\bibitem{HS:GHLS}
M.~Harada and C.~Sasaki,
Phys.\ Rev.\  D {\bf 73}, 036001 (2006).

\bibitem{rest}
See, e.g.,
V.~Bernard and U.~G.~Meissner,
Nucl.\ Phys.\ A {\bf 489}, 647 (1988);
T.~Hatsuda and T.~Kunihiro,
Phys.\ Rept.\  {\bf 247}, 221 (1994)
[hep-ph/9401310];
R.~D.~Pisarski,
hep-ph/9503330;
R.~Rapp and J.~Wambach,
Adv.\ Nucl.\ Phys.\  {\bf 25}, 1 (2000);
F.~Wilczek,
hep-ph/0003183;
G.~E.~Brown and M.~Rho,
Phys.\ Rept.\  {\bf 363}, 85 (2002).

\bibitem{BR-scaling}
G.~E.~Brown and M.~Rho,
Phys.\ Rev.\ Lett.\  {\bf 66}, 2720 (1991).

\bibitem{HY:VM}
M.~Harada and K.~Yamawaki,
Phys.\ Rev.\ Lett.\  {\bf 86}, 757 (2001).

\bibitem{HY:PRep}
M.~Harada and K.~Yamawaki,
Phys.\ Rept.\  {\bf 381}, 1 (2003).

\bibitem{HS:VM}
M.~Harada and C.~Sasaki,
Phys.\ Lett.\ B {\bf 537}, 280 (2002).

\bibitem{HKR:VM}
M.~Harada, Y.~Kim and M.~Rho,
Phys.\ Rev.\ D {\bf 66}, 016003 (2002).

\bibitem{BKUYY}
M.~Bando, T.~Kugo, S.~Uehara, K.~Yamawaki and T.~Yanagida,
Phys.\ Rev.\ Lett.\  {\bf 54}, 1215 (1985):

\bibitem{BKY:PRep}
M.~Bando, T.~Kugo and K.~Yamawaki,
  Phys.\ Rept.\  {\bf 164}, 217 (1988).

\bibitem{ceres}
G.~Agakishiev {\it et al.} [CERES Collaboration],
Phys.\,Rev.\,Lett.\,{\bf 75}, 1272 (1995).

\bibitem{LKB}
G.~Q.~Li, C.~M.~Ko and G.~E.~Brown,
Phys.\ Rev.\ Lett.  {\bf 75}, 4007 (1995).

\bibitem{KEK-PS}
K.~Ozawa {\it et al.}  [E325 Collaboration],
Phys.\ Rev.\ Lett.\  {\bf 86}, 5019 (2001);
  M.~Naruki {\it et al.},
  Phys.\ Rev.\ Lett.\  {\bf 96}, 092301 (2006).

\bibitem{trnka}
D.~Trnka\,{\it et al.}\,[CBELSA/TAPS Collaboration],
Phys.\,Rev.\,Lett.\,{\bf 94}, 192303 (2005).

\bibitem{SB:STAR}
E.~V.~Shuryak and G.~E.~Brown,
Nucl.\ Phys.\ A {\bf 717}, 322 (2003).

\bibitem{NA60}
  S.~Damjanovic {\it et al.}  [NA60 Collaboration],
  J.\ Phys.\ G {\bf 31}, S903 (2005);

\bibitem{NA60:2}
S.~Damjanovic {\it et al.}  [NA60 Collaboration],
Nucl.\ Phys.\  A {\bf 774}, 715 (2006).

\bibitem{Brown:2005ka-kb}
  G.~E.~Brown and M.~Rho,
  arXiv:nucl-th/0509001;
  arXiv:nucl-th/0509002.

\bibitem{HR}
H.~van Hees and R.~Rapp,
  arXiv:hep-ph/0604269.

\bibitem{SG}
B.~Schenke and C.~Greiner,
Phys.\ Rev.\ Lett.\  {\bf 98}, 022301 (2007).

\bibitem{HS:VD}
M.~Harada and C.~Sasaki,
Nucl.\ Phys.\ A {\bf 736}, 300 (2004).

\bibitem{HS:dilepton}
M.~Harada and C.~Sasaki,
Phys.\ Rev.\  D {\bf 74}, 114006 (2006).

\bibitem{BLR}
G.~E.~Brown, C.~H.~Lee and M.~Rho,
arXiv:nucl-th/0507073.

\bibitem{rhoA1model}
See, e.g.,
C.~Song and V.~Koch,
Phys.\ Lett.\ B {\bf 404}, 1 (1997);
M.~Urban, M.~Buballa and J.~Wambach,
Phys.\ Rev.\ Lett.\  {\bf 88}, 042002 (2002).

\bibitem{BKY:NPB}
M.~Bando, T.~Kugo and K.~Yamawaki,
Nucl.\ Phys.\ B {\bf 259}, 493 (1985).

\bibitem{BFY:GHLS}
  M.~Bando, T.~Fujiwara and K.~Yamawaki,
  Prog.\ Theor.\ Phys.\  {\bf 79}, 1140 (1988).

\bibitem{Kaiser:1990yf}
N.~Kaiser and U.~G.~Meissner,
Nucl.\ Phys.\ A {\bf 519}, 671 (1990).

\bibitem{ChPT}
S.~Weinberg,
Physica A {\bf 96}, 327 (1979);
J.~Gasser and H.~Leutwyler,
Annals Phys.\  {\bf 158}, 142 (1984);
Nucl.\ Phys.\ B {\bf 250}, 465 (1985).

\bibitem{SVZ}
M.~A.~Shifman, A.~I.~Vainshtein and V.~I.~Zakharov,
Nucl.\ Phys.\ B {\bf 147}, 385 (1979);
Nucl.\ Phys.\ B {\bf 147}, 448 (1979).

\bibitem{Weinberg}
  S.~Weinberg,
  Phys.\ Rev.\ Lett.\  {\bf 18}, 507 (1967).


\end{thebibliography}
\end{document}